\documentclass
[%
preprint,
 showpacs,preprintnumbers,
 amsmath,amssymb,
 aps,
 pre,
 floatfix,
 longbibliography
]{revtex4-1}

\usepackage{booktabs}
\usepackage{array}
\newcolumntype{L}[1]{>{\raggedleft\arraybackslash}p{#1}}
\AtBeginDocument{
\heavyrulewidth=.08em
\lightrulewidth=.05em
\cmidrulewidth=.03em
\belowrulesep=.65ex
\belowbottomsep=0pt
\aboverulesep=.4ex
\abovetopsep=0pt
\cmidrulesep=\doublerulesep
\cmidrulekern=.5em
\defaultaddspace=.5em
}

\usepackage{color}
\usepackage{xcolor}
\usepackage{graphicx}% Include figure files
\usepackage{dcolumn}% Align table columns on decimal point
\usepackage{bm}% bold math
\usepackage{cancel}
\usepackage[utf8]{inputenc}

\begin{document}

%\preprint{APS/123-QED}

\title{Self-diffusion coefficient as a function of the thermodynamic factor}

\author{M. Sampayo Puelles}
\author{M. Hoyuelos}
\email{hoyuelos@mdp.edu.ar}

\affiliation{Instituto de Investigaciones F\'isicas de Mar del Plata (IFIMAR -- CONICET), Departamento de F\'isica, Facultad de Ciencias Exactas y Naturales,
Universidad Nacional de Mar del Plata, De\'an Funes 3350, 7600 Mar del Plata, Argentina}

\date{\today}% It is always \today, today,
             % but any date may be explicitly specified

\begin{abstract}
Much effort has been put into developing theories for dense fluids, as a result of these efforts many theories work for a certain type of particle or in a certain concentration regime. Rosenfeld proposed a dependence of the self-diffusion coefficient on the excess entropy. Our proposal is similar to Rosenfeld’s in that it also
attempts to describe diffusion in terms of a thermodynamic function but, instead of the excess entropy, we use the thermodynamic factor, or the excess chemical potential. Simulations were taken for hard spheres and our model was fitted with two free parameters. Simulations were then carried out for a Lennard Jones gas and our model correctly described the new data with the value of the free parameters that we had obtained for hard spheres.  This is a feature of our model that we wish to emphasize, since the usual situation is that parameters have to be re-adjusted
for different interaction potentials. An experimental xenon self-diffusion data set was used as an example where the model can be applied, especially in the high-density regime.

\end{abstract}

\maketitle

Since the development of the Boltzmann theory, and the Chapman-Enskog method \cite{chapman}, that permits the evaluation of transport coefficients in a dilute gas, many efforts have been devoted to extend the theory to dense fluids. An early attempt was the Enskog theory for a dense gas of hard spheres (\cite{enskog}, see also \cite[Sec.\ 9.3.1]{silva}); it reproduces transport coefficients for moderate concentrations, in the so-called Eskong regime (specially accurate for viscosity and heat conductivity for reduced concentrations up to 0.4 or 0.5). Considering that the transport mechanisms in real fluids do not essentially differ from those in hard spheres, the Modified Enskog Theory extends the previous results to real gases \cite{hanley}. Other approaches involve an effective hard sphere diameter \cite{barker,wca,andersen,verlet,lado} (the softness of a repulsive potential is accounted for by an effective diameter that depends on temperature and possibly also on density, so that the properties of a fluid can be calculated by the corresponding hard sphere model), functions of the free volume \cite{dymond,hildebrand,batschinski,doolittle,cohen2,macedo} or of the excess entropy \cite{rosenfeld,rosenfeld2,dzugutov}. Perturbation theories are based on the separation of the repulsive and attractive effects of the intermolecular interaction potential. The Hard Sphere model is often used to describe the repulsive part and the effect of the attractive forces is regarded as a perturbation.  Hard spheres with a temperature dependent size, combined with the van der Waals theory, have been also used to reproduce transport properties of real fluids \cite{dymond2,dymond3,dymond4}.

Rosenfeld \cite{rosenfeld,rosenfeld2} studied the effect of attractive forces by comparing results for purely repulsive potentials and Lennard-Jones (LJ) systems, and found that the self diffusion coefficient against excess entropy on logarithmic scale lies on nearly the same line (see also \cite[Sec.\ 9.3.7]{silva}). The exponential dependence on the excess entropy was verified for several substances at moderate and large concentrations \cite{dzugutov}. Nevertheless, Rosenfeld observed that this exponential dependence can not be extended to the small concentration regime \cite{rosenfeld3}.

The results of these efforts is a set of theories that hold for specific kind of particles or for different concentration regimes, and that usually require the adjustment of free parameters.

Our proposal is similar to Rosenfeld's in that it also attempts to describe diffusion in terms of a thermodynamic function but, instead of the excess entropy, we use the thermodynamic factor, or the excess chemical potential. The proposal is based on two recent results. The first one applies to a general particle system divided into small cells; each cell is in local thermal equilibrium, interactions between cells are neglected. Interactions at the macroscopic level are represented by the excess chemical potential $\mu_\text{ex}$. It has been shown \cite{dimuro} that the average transition rate between two neighboring cells with $n_1$ and $n_2$ particles is given by
\begin{equation}\label{e.W}
	W_{n_1,n_2} = \nu \frac{e^{-\beta (\mu_{\text{ex},n_2}-\mu_{\text{ex},n_1})/2}}{\sqrt{\Gamma_{n_1}\Gamma_{n_2}}},
\end{equation}
where the index order in $W_{n_1,n_2}$ indicates that the jump is from cell 1 to cell 2,  $\Gamma_{n_i} = 1 + \beta n_i \frac{\partial \mu_{\text{ex},n_1}}{\partial n_i}$ is the thermodynamic factor, and $\nu$ is the jump attempt frequency that depends on the features of the substratum; $\beta = 1/(k_B T)$ with $T$ the temperature and $k_B$ the Boltzmann constant. Eq.\ \eqref{e.W} reproduces the Darken equation \cite{darken} (see \cite{dimuro}) that gives a relationship between the self-diffusion and the collective diffusion coefficients; it also has been generalized to quantum systems of non-interacting particles \cite{hoyuelos1}. Since $W_{n_1,n_2}$ is an average transition rate for any of the $n_1$ particles in cell 1, it does not provide the transition rate of a tagged particle, necessary to calculate the self-diffusion coefficient. Nevertheless, it suggests that, if the self-diffusion coefficient depends on a thermodynamic function, this function can be the thermodynamic factor or the excess chemical potential.

The second result is the numerical verification that, in fact, $D/D_0$ is a thermodynamic function, where $D$ is the self-diffusion coefficient and $D_0$ is its value at small concentration. This has been verified for pseudo-hard spheres and for the Lennard-Jones potential, with a Langevin thermostat in both cases \cite{marchioni}.

Here, we propose a specific form for this thermodynamic function, with two free parameters, that has the following features. It correctly describes numerical results of $D/D_0$ for hard spheres (HS) for the whole concentration range. The same parameters fitted for HS are also valid to approximately represent numerical results for the Lennard-Jones (LJ) potential. 

The paper is organized as follows. The model is introduced in Sec. \ref{s.model}. Our model fit for HS and the simulations and model fit for LJ gas are included in Sec. \ref{s.results}. The example of xenon experimental data and our model fitting are included in Sec. \ref{s.expdata}. Conclusions are presented in Sec. \ref{s.conclusions}.

\section{The model}
\label{s.model}

We introduce a general expression for the self-diffusion coefficient $D$. Numerical results of $D$ for hards spheres and for the Lennard-Jones potential are used in the next section in order to verify the validity of the proposal. As mentioned in the introduction, in a previous work \cite{dimuro} we were able to obtain an expression for the transition rate adopting a coarse grained picture, where microscopic details are lost. The system is divided into cells, each cell contains many particles, and it can be demonstrated that the transition rate per particle between cells depends on the excess chemical potential, $\mu_\text{ex}$, and the thermodynamic factor, $\Gamma$. We propose the following expression for the self-diffusion coefficient:
\begin{equation}\label{e.DD0}
	D/D_0 = \exp[a (\Gamma - 1) + b\, \rho\, \sigma_\text{HS}^3],
\end{equation}
where $a$ and $b$ are parameters to be adjusted; the number density is $\rho=n/V$ and $\rho\, \sigma_\text{HS}^3$ is a dimensionless number density, with $\sigma_\text{HS}$ equal to the particle diameter for hard spheres (HS), or an equivalent hard-sphere diameter in other cases, and $n$ and $V$ are particle number and volume respectively. In the limit of small concentration we have that $\Gamma\rightarrow 1$ and $D\rightarrow D_0$.  

Let us analyze the model for a repulsive interaction potential. Self-diffusivity should decrease as concentration increases due to system clogging. The term $b\rho\, \sigma_\text{HS}^3$ in the exponential can be interpreted as a simplification of a free-volume ($V_f$) theory with exponential dependence on $V_f$ for small concentration \cite{cohen2,liu}, with $b$ negative. This exponential decay with density is not enough for an appropriate description. The information regarding specific features of the interaction is represented by the term that includes the thermodynamic factor in the exponential in \eqref{e.DD0}: $a(\Gamma - 1)$. The excess chemical potential increases for increasing concentration, since interaction is repulsive, and $\Gamma = 1 + \beta \rho \frac{\partial \mu_\text{ex}}{\partial \rho}$ becomes greater than 1. In this case, parameter $a$ is negative. The motivation for including the thermodynamic factor is, as mentioned in the introduction, the dependence of transition probabilities on $\Gamma$, see Eq.\  \eqref{e.W}. It is expected that the combination of both terms provides a satisfactory description of self-diffusivity in the whole range of concentration.
Eq.\ \eqref{e.DD0} is equivalent to
\begin{equation}\label{e.DD02}
	D/D_0 = \exp(a \beta \rho\, \mu_\text{ex}' +  b\, \rho\, \sigma_\text{HS}^3),
\end{equation}
where $\mu_\text{ex}' = \frac{\partial \mu_\text{ex}}{\partial \rho}$

The information needed to calculate the thermodynamic factor, or the excess chemical potential, is provided by the equation of state (EOS), usually represented by the compressibility factor, $Z = \beta p/\rho$, where $p$ is the pressure. The thermodynamic factor is given by (see Appendix A)
\begin{equation}
\Gamma = Z + \rho \frac{\partial Z}{\partial \rho}.
\end{equation}

The Carnahan-Starling EOS is frequently used due to its simplicity and sufficient accuracy to derive thermodynamic properties of the fluid HS system \cite{carnahan}. Its analytical expression is
\begin{equation} \label{e.CS}
	Z=\frac{1 + \varphi + \varphi^2 - \varphi^3}{(1-\varphi)^3},
\end{equation} 
where $\varphi = \rho \sigma_\text{HS}^3 \pi/6$ is the packing fraction.

%The hard sphere and attractive parts of the chemical potential are obtained from the corresponding separation of the compressibility factor: $Z=Z_{\mathrm{HS}} + Z_{\mathrm{attr}}$

For the Lennard-Jones system, we use the EOS obtained by Pieprzyk \textit{et al.}\ \cite{piep_eqS}, see Appendix B. The results derived from this EOS for the thermodynamic factor are similar to the ones that can be calculated from other approximate equations for the Lennard-Jones system as, for example, the EOS of Ree \cite{ree}, Kolafa and Nezbeda \cite{kolafa} and Mecke \cite{mecke}.

The self-diffusion coefficient at small concentration is calculated from the kinetic theory formula based upon the Boltzmann equation:
\begin{equation} 
D_0=\frac{3}{8\rho\sigma^2} \left(\frac{k_B T}{\pi m}\right)^{1/2} \frac{1}{\Omega^{(1,1)*}},
\end{equation}
where $\Omega^{(s,l)*}$ are dimensionless collision integrals obtained by dividing them by their corresponding HS values, and $\sigma$ is a characteristic distance parameter between colliding molecules. For the HS system, $\sigma$ is the sphere diameter and  $\Omega^{(s,l)*}$ becomes unitary.

%Enskog developed the first successful theory of dense gases onto the previous dilute gas model, taking into account that the diameter of a molecule is no longer negligible compared with the interparticle distance.
%Concerning the number of binary collisions that occur in a gas, its frequency was modified by a factor $g(\sigma)$: the radial distribution function at contact, which is unity at low density.
%The resultant Enskog equation for the diffusion coefficient of dense HS fluid, subscript E, is
%\begin{equation}
%\frac{\rho D_E}{\rho_o D_o}=\frac{1}{g(\sigma)}
%\end{equation}
%
%The value of $g(\sigma)$ can be obtained from the equation of state (EOS) as follows:
%\begin{equation}
%\frac{P}{\rho k_B T}=1+b\rho g(\sigma)
%\end{equation}
%where b is the second virial coefficient (often set equal to the van der Waals excluded volume) which for HS is defined as 
%\begin{equation}
%b=2\pi \sigma^3 /3 .
%\end{equation}
%
%By using the HS virial expansion
%\begin{equation}
%\frac{P}{\rho k_B T}=1+b\rho + 0.625 (b\rho)^2 + 0.2869 (b\rho)^3 + 0.115 (b\rho)^4 + ... 
%\end{equation} 
%one easily finds
%\begin{equation}
%g(\sigma)=1 + 0.625b\rho + 0.2869 (b\rho)^2 + 0.115 (b\rho)^3 + ...
%\end{equation}
%
%
%
%and then 
%\begin{equation}
%g(\sigma)=\frac{1-\varphi/2}{(1-\varphi)^3}
%\end{equation}
%where $\varphi$ is the HS packing fraction, which is given by 
%\begin{equation}
%\varphi=\frac{\pi}{6} \rho \sigma^3 \equiv \frac{\pi \rho^*}{6}
%\end{equation}
%$\rho^* \equiv \rho \sigma^3$ being the reduced number density.

\begin{figure} %[ht]
	\centering
	\includegraphics[width=\linewidth]{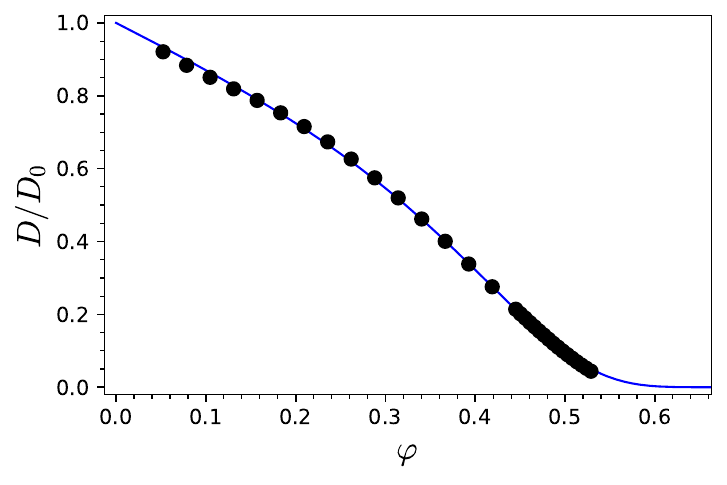}
	\caption{Data fitting of the self-diffusion coefficient, relative to the small concentration value, $D/D_0$, against the packing fraction, $\varphi$, for hard spheres. The simulation data was taken from Pieprzyk \cite{piep}. The curve corresponds to Eq.\ \eqref{e.DD0} with $ a=-0.0336$ and $b=-0.958$ (and with $\rho \sigma_\text{HS}^3 = \varphi\, 6/\pi$). } \label{HS}
\end{figure}

\section{Results}
\label{s.results}

For the HS system, we used the numerical results obtained by Pieprzyk \cite{piep}. The parameters $a$ and $b$ of the model given by Eq.\ \eqref{e.DD0} were adjusted using the least squares method. The thermodynamic factor, $\Gamma$, was calculated using the Carnahan-Starling EOS, Eq.\ \eqref{e.CS}. The values obtained for the parameters are:
\begin{equation}\label{e.param}
	 a =-0.0336, \qquad b=-0.958.
\end{equation}
Fig.\ \ref{HS} shows that the model successfully represents the data with an accuracy similar to the dot size.

\begin{figure}%[h!]
	\centering
	\includegraphics[width=\linewidth]{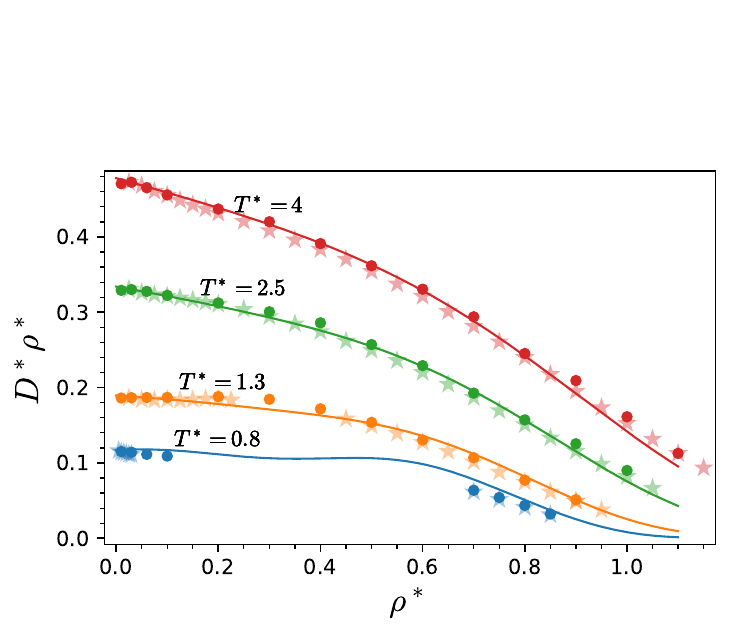}
	\caption{Our simulations are shown in dark points compared to those of Meier in lighter stars.} \label{simu_vs_data}
\end{figure}

For the Lennard-Jones fluid, simulations were performed for the diffusion coefficient  for four different temperatures. Fig.\ \ref{simu_vs_data} shows the results of our simulations in dark dots for $D^*\rho^*$ against $\rho^*$ for $T^*=0.8$, 1.3, 2.5 and 4; simulation results of Maier \cite{meier} in lighter stars are also included for comparison. The asterisk superscript indicates dimensionless quantities; for the LJ system they are defined as 
\begin{equation}\label{e.LJadim}
	 \rho^*= \rho \sigma^3, \quad T^*=\frac{k_B T}{\epsilon}, \quad D^*=D \frac{1}{\sigma} \sqrt{\frac{m}{\epsilon}},
\end{equation}
where $\epsilon$ is the depth of the well of the LJ potential, $m$ is the mass of the particles, $\sigma$ is given by the distance at which the interaction potential is equal to zero.
Lammps software \cite{plimpton} was used for the molecular dynamics simulations. The parameters used for the simulations are the following. Number of particles: 3200; thermalization time with a Langevin thermostat: 50000 steps; time step 0.001; data gathering time: 100000 steps; cutoff radius for the Lennard-Jones interaction: 2.5. 
The Green Kubo formula was used, for the different temperatures over the entire density range, to calculate the diffusion coefficient as an integral of the velocity auto-correlation function:  
\begin{equation}
D= \frac{1}{3}  \int_{0}^{\infty} \langle \textbf{v}_i(0) \cdot \textbf{v}_i(t) \rangle dt.
\end{equation}

Let us notice that, for interactions other than hard spheres, we have to specify an effective hard-sphere diameter, $\sigma_\text{HS}$, in order to evaluate Eq.\ \eqref{e.DD0}, or \eqref{e.DD02}. There are several options. The form of the radial distribution function is primarily determined by repulsive forces, while attractive interactions play a secondary role. This promoted the development of the perturbation approaches for dense fluids, which usually combine hard spheres as an first approximation, for the major excluded volume and packing effects, with an effective diameter dependent on temperature and possibly on density. The essential assumption of the effective hard sphere diameter (EHSD) method is that the properties of a fluid can be calculated by the corresponding HS model, if the molecular diameter is replaced by an EHSD, $\sigma_\text{HS}$. Various EHSD equations have been proposed in the literature. Barker and Henderson published the first successful analysis, adopting the HS and LJ systems as the unperturbed and perturbing potentials; this EHSD is only temperature dependent. Another milestone in the perturbation theory is the work of Weeks, Chandler and Andersen who splitted up the LJ potential into a reference part containing all repulsive forces, and a perturbing part containing all forces of attraction. The WCA perturbation theory yields an EHSD dependent on both temperature and density. We use an expression only temperature dependent, other expressions from the literature were tested with similar results. 
The following expression for the effective hard sphere diameter, proposed in Ref.\ \cite{speedy}, was used:
\begin{equation}
	\frac{\sigma_\text{HS}}{\sigma} = 2^{\frac{1}{6}} \left[1+(2T^*)^{\frac{1}{2}}\right]^{-\frac{1}{6}};
\end{equation}
see also Eq.\ (9.55) in Ref.\ \cite{silva}.

Figure \ref{simu_vs_data} shows a good agreement between simulations results and the model of Eq.\ \eqref{e.DD0} with parameters $a$ and $b$ given by \eqref{e.param}, that is, the same parameters used for the fitting of the HS results. 

%\begin{figure}%[hbt]
%	\centering
%	\includegraphics[width=0.85\linewidth]{simu_y_fiteo_T955}
%	\caption{Simulations are in dots and the curve represent our model.} \label{simu_y_fiteo_T955}
%\end{figure}

%\begin{figure}[hbt]
%	\includegraphics[width=0.85\linewidth]{cuatroG_955}
%	\caption{Simulation and model} \label{cuatroG_955}
%\end{figure}

\section{Experimental data}
\label{s.expdata}

We use self-diffusion experimental data of Peereboom \textit{et al.} \cite{peereboom0,peereboom}  for the $^{129}\mathrm{Xe}$ isotope for four different temperatures: $248\, K$, $273\, K$, $298\, K$ and $343\, K$; the first two are below the critical temperature $T_c = 290\, K$. A description of the xenon system based on the LJ model is used. The parameters $\sigma=0.3924\ \mathrm{nm}$ and $\epsilon/k_B=257.4\, K$, reported in \cite{peereboom}, were used to go to reduced Lennard-Jones units.

\begin{figure}%[hbt]
	\centering
	\includegraphics[width=\linewidth]{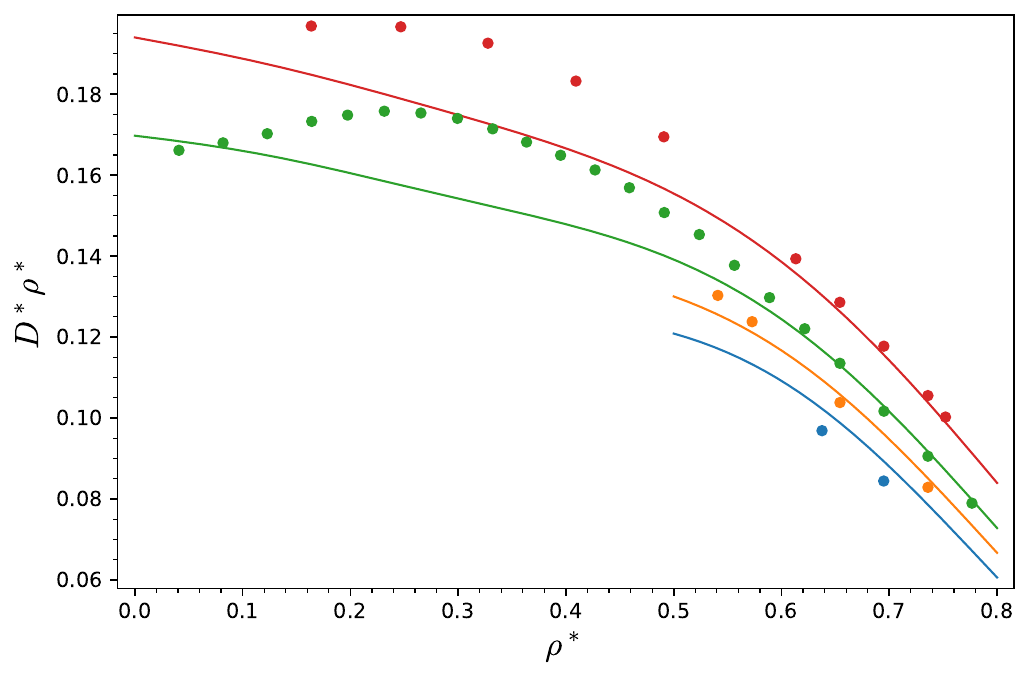}
	\caption{Experimental results, taken from \cite{peereboom}, for the scaled self-diffusion times density, $D^* \rho^*$, against density, $\rho^*$, for temperatures $T=$ $248\, K$ (blue dots), $273\, K$ (orange), $298\, K$ (green) and $343\, K$ (red). The curves correspond to the model of Eq.\ \eqref{e.DD0}, where the Lennard-Jones EOS was used.} \label{simu_y_expdata}
\end{figure}

Fig.\ \ref{simu_y_expdata} shows the mentioned experimental data and also the theoretical curve from Eq.\ \eqref{e.DD0}, for $D^*\rho^*$ against $\rho^*$. It is known that there are some discrepancies when the Lennard-Jones model is used to describe xenon \cite[p.\ 114]{Meier2002}. The most remarkable is that the isotherm close and above the critical temperature ($T=298\, K$, green dots in Fig. \ref{simu_y_expdata}) shows a maximum near the critical density ($\rho_c = 1120.8$ kg/m$^3$, or $\rho_c^* =0.32$), while isotherms in the Lennard-Jones model do not have such pronounced maxima (see Fig.\ \ref{simu_vs_data}). As a consequence, our model also shows that discrepancy in the same region. On the other hand, there is a good agreement between model and experiments for large density, in the liquid region. Parameters $a$ and $b$ are, again, the ones obtained from the HS system \eqref{e.param}.

%\begin{figure}[hbt]
%	\includegraphics[width=0.85\linewidth]{cuatroG_expdata}
%	\caption{Experimental data and model} \label{cuatroG_expdat}
%\end{figure}

\section{Conclusions}
\label{s.conclusions}

In a dynamic mean field approximation, where it is assumed that the transition rate for one particle is approximately equal to the average transition rate of the particles in a cell, we can obtain, using \eqref{e.W}, that the self-diffusion coefficient is $D = D_0/\Gamma$ and the collective diffusion coefficient is $D_c = D_0$ \cite{dimuro}; combining both results we have the Darken relationship $D = D_c/\Gamma$, that has been successfully applied to diffusion in solids \cite{mehrer} but, in general, this approximation does not hold for dense fluids. In the derivation of Eq.\ \eqref{e.W} it was assumed that cells are large enough to be taken as thermodynamic systems, and the excess chemical potential is evaluated at the thermodynamic limit. Further studies are required to verify this assumption, since small size systems introduce corrections to the chemical potential that may be relevant to the diffusion process.

There are several factors to be taken into account in order to arrive to an accurate and general description of diffusion in dense systems. At large densities, an important correlation effect is back-scattering, whereby a sphere closely surrounded by a shell of neighboring spheres becomes increasingly locked in and reverses its velocity on collision, which decreases diffusion. On the other hand, Alder and Wainwright \cite{alder3} showed that the velocity autocorrelation function has a slower than exponential decay (long time tails), generated by the presence of vortices, that, instead, enhance diffusion. If the hypothesis (partially verified in Ref.\ \cite{marchioni}) that $D/D_0$ is a thermodynamic function is correct, then the information concerning the influence of correlations on diffusion is already contained in the EOS. This is a result that represents an important simplification for the description of diffusion in dense fluids.

In this paper, we propose a specific form for that thermodynamic function with two adjustable parameters. The model \eqref{e.DD0} includes an exponential function of the thermodynamic factor that is motivated by the form of the transition rates \eqref{e.W}; the exponent includes also a term proportional to the density. The model provides an accurate description of the self-diffusion for hard spheres in the whole density range. The same parameters, $a$ and $b$, that were used to fit the model to the hard sphere data, were  used for the Lennard-Jones potential, giving also a satisfactory description of the numerical data. This is a feature of our model that we wish to emphasize, since the usual situation is that parameters have to be re-adjusted for different interaction potentials.

A set of experimental data of self-diffusion in xenon was used as an example were the model \eqref{e.DD0} can be applied. The EOS for the Lennard-Jones potential was used to describe xenon, but it is known that this description is not completely accurate, specially close to the critical point. Nevertheless, the model approximately matches the experimental results for large densities.

\begin{acknowledgments}
This work was partially supported by Consejo Nacional de Investigaciones Cient\'ificas y T\'ecnicas (CONICET, Argentina, PUE 22920200100016CO).
\end{acknowledgments}

\section*{Appendix A}

In this appendix we derive the relationship between EOS and thermodynamic factor [see, for example, Eqs.\ (9) and (11) in \cite{kolafa}]. 
Let us consider a cell of volume  $\mathcal{V}$, number of particles $n$ and temperature $T$. If $F$ is the free energy, the pressure is $p=-\left.\frac{\partial F}{\partial \mathcal{V}}\right|_{n,T}$ and the chemical potential is $\mu = \left. \frac{\partial F}{\partial n}\right|_{\mathcal{V},T}$. The free energy per particle is $f= F/n$, a function of density $\rho=n/\mathcal{V}$ and temperature. The pressure and chemical potential are $ p=\rho^2 \frac{\partial f}{\partial \rho}$ and $\mu= f+p/\rho$. Knowing that the compressibility factor is $Z=\beta p /\rho$, the thermodynamic factor can be written as 
\begin{equation}
	\begin{split}
		\Gamma & =\beta\rho\frac{\partial \mu}{\partial \rho} = \beta \rho \left(\frac{\partial f}{\partial \rho} + \frac{\partial (p/\rho)}{\partial\rho}\right) \\
		&=\beta p/\rho + \rho \frac{\partial(\beta p /\rho)}{\partial \rho} \\
		&= Z + \rho\frac{\partial Z}{\partial \rho}.
	\end{split}
\end{equation}
%Insted of making the replacement in the third kine, it can be simplified to $$\Gamma=\beta \frac{\partial p}{\partial \rho},$$ giving an alternative expression for the thermodynamic factor that is connected to the isothermal compressibility.
 
%For residual quantities $\mu_{ex}=\mu-\mu_{id}$ and $f_{ex}= f-f_{id}$, knowing that $f_{id}=\mu_{id}-\beta^{-1}$ and $\mu_{id}=\mu^o + \beta^{-1} \mathrm{ln} n$, we have
%\begin{equation}
%	p=\rho^2 \frac{\partial f_{ex}}{\partial \rho} +\rho \beta^{-1},
%\end{equation}
%\begin{equation}
%	\mu_{ex}= f_{ex} + p/\rho - \beta^{-1}.
%\end{equation}
%
%Rewriting the previous expressions in terms of the compressibility factor, we get 
%\begin{equation}
%	\beta \frac{\partial f_{ex}}{\partial\rho} = (Z-1)/\rho,
%\end{equation}
%\begin{equation}
%	\beta \mu_{ex}=\beta f_{ex}+Z-1.
%\end{equation}
%combaining the last two equations, we obtain eq. \ref{eq_mu}.

\section*{Appendix B}

For completeness, we reproduce here the EOS of Pieprzyk \textit{et al.} for the Lennard-Jones potential \cite{piep_eqS}:
\begin{equation}
	\begin{split}
	&P(\rho, T) = \rho T +\rho^2 \left(x_1T+x_2\sqrt{T}+x_3+\frac{x_4}{T}+\frac{x_5}{T^2}\right)\\
&\quad + \rho^3 \left(x_6T+x_7+\frac{x_8}{T}+\frac{x_9}{T^2}\right) + \rho^4 \left(x_{10}T+x_{11}+\frac{x_{12}}{T}\right)\\
&\quad +\rho^5x_{13}+\rho^6\left(\frac{x_{14}}{T}+\frac{x_{15}}{T^2}\right)+\rho^7  \frac{x_{16}}{T}+\rho^8\left(\frac{x_{17}}{T}+\frac{x_{18}}{T^2}\right) \\
&\quad +\rho^9 \left(\frac{x_{19}}{T^2}\right) + \left[\rho^3 \left(\frac{x_{20}}{T^2}+\frac{x_{21}}{T^3}\right) + \rho^5\left(\frac{x_{22}}{T^2}+\frac{x_{23}}{T^4}\right)\right.\\
&\quad  + \left.\rho^7\left(\frac{x_{24}}{T^2}+\frac{x_{25}}{T^3}\right)+\rho^9\left(\frac{x_{26}}{T^2}+ \frac{x_{27}}{T^4}\right) \right.\\ 
&\quad  + \left.\rho^{11}\left(\frac{x_{28}}{T^2}+\frac{x_{29}}{T^3}\right)+\rho^{13}\left(\frac{x_{30}}{T^2}+\frac{x_{31}}{T^3}+\frac{x_{32}}{T^4}\right)\right] e^{-3\rho^2},
	\end{split}
\end{equation}
where $\rho$ and $T$ are scaled dimensionless quantities in which the asterisk was omitted to lighten the notation. See Table I in Ref.\ \cite{piep_eqS} for the values of parameters $x_i$, $i=1,\cdots,32$.

\begin{multline*}
\end{multline*}

\bibliography{HSandLJ.bib}

\end{document}